# Influence of direct deposition of dielectric materials on the optical response of monolayer WS$_2$

Tinghe Yun[1,2,3], Matthias Wurdack[1,3], Maciej Pieczarka[1,4], Semonti Bhattacharyya[1,5], Qingdong Ou[1,2], Christian Notthoff[3], Patrick Kluth[3], Michael S. Fuhrer[1,5], Andrew G. Truscott[3], Eliezer Estrecho[1,3], and Elena A. Ostrovskaya [1,3]

[1]ARC Centre of Excellence in Future Low-Energy Electronics Technologies
[2]Department of Materials Science and Engineering, Monash University, Clayton, Victoria, 3800, Australia
[3]Research School of Physics, The Australian National University, Canberra, ACT 2601, Australia
[4]Department of Experimental Physics, Wrocław University of Science and Technology, Wyb. Wyspiańskiego 27, Wrocław 50-370, Poland
[5]School of Physics and Astronomy, Monash University, Clayton, Victoria, 3800, Australia

**Abstract**

The integration of two-dimensional transition metal dichalcogenide crystals (TMDCs) into a dielectric environment is critical for optoelectronic and photonic device applications. Here, we investigate the effects of direct deposition of different dielectric materials (Al$_2$O$_3$, SiO$_2$, SiN$_x$) onto atomically thin (monolayer) TMDC WS$_2$ on its optical response. Atomic layer deposition (ALD), electron beam evaporation (EBE), plasma enhanced chemical vapour deposition (PECVD), and magnetron sputtering methods of material deposition are investigated. The photoluminescence (PL) measurements reveal quenching of the excitonic emission after all deposition processes. The reduction in neutral exciton PL is linked to the increased level of charge doping and associated rise of the trion emission, and/or the localized (bound) exciton emission. Furthermore, Raman spectroscopy allows us to clearly correlate the observed changes of excitonic emission with the increased levels of lattice disorder and defects. Overall, the EBE process results in the lowest level of doping and defect densities and preserves the spectral weight of the exciton emission in the PL, as well as the exciton oscillator strength. Encapsulation with ALD appears to cause chemical changes, which makes it distinct from all other techniques. Sputtering is revealed as the most aggressive deposition method for WS$_2$, fully quenching its optical response. Our results demonstrate and quantify the effects of direct deposition of dielectric materials onto monolayer WS$_2$, which can provide a valuable guidance for the efforts to integrate monolayer TMDCs into functional optoelectronic devices.

Two-dimensional (2D) transition metal dichalcogenide crystals (TMDCs) have been attracting a lot of interest in the past decade due to their unique optical and electronic properties [1-6]. These 2D semiconductors exhibit a layered structure with a strong in-plane covalent bonding and weak out-of-plane van der Waals forces [7]. The single-layer 2D (monolayer) TMDCs have a direct bandgap and show strong photoluminescence (PL) emission produced by radiative recombination of excitons and excitonic complexes formed upon photoexcitation [8]. The reduced dielectric screening effect at the monolayer limit results in robust light-matter interactions [9, 10]. Therefore, monolayer TMDCs are recognized as promising semiconductors for optoelectronics and photonics devices, such as transistors [6, 11, 12] and light-emitting diodes (LEDs) [13], as well as for fundamental studies of quantum behaviour of excitons [14]. The integration of monolayer TMDCs into a dielectric environment is a crucial technological step for future device applications.

Scalable fabrication steps for transistors or LEDs based on TMDCs typically include the deposition of dielectric materials with either high or low dielectric constants, $\kappa$, onto the fragile monolayers. For example, the top-gate of a field-effect transistor should be a high-$\kappa$ dielectric [15]. The direct deposition of a dielectric on top of a monolayer TMDC can be performed by different techniques. The atomic layer deposition (ALD) is the most commonly used method due to its high-quality thin film finish [16-18]. However, for many optoelectronic applications it is critical that the deposition method preserves the optical properties of the monolayer. It has been reported that the excitonic PL emission of the monolayer $MoS_2$, $MoSe_2$, $WS_2$, and $WSe_2$ is quenched after the deposition of $Al_2O_3$ (high-$\kappa$ dielectric) via the ALD [19], but very little is known of the effect of other deposition techniques and lower-$\kappa$ dielectric materials. In this work, we focus on monolayer $WS_2$, which has pronounced exciton photoluminescence at room temperature. In particular, we investigate the change of the optical response of the monolayer, after depositing either a high-$\kappa$ or a low-$\kappa$ dielectric material by one of the four commonly used thin film deposition techniques: atomic layer deposition (ALD), electron beam evaporation (EBE), plasma enhanced chemical vapour deposition (PECVD), and radio-frequency (RF) magnetron sputtering. We probe the photoluminescence spectra, reflectivity spectra and Raman scattering spectra to investigate the changes in excitonic properties and material structure. All of the above techniques lead to strong excitonic PL quenching caused by the exposure to either high-energy particles or chemical reactions during the dielectric deposition process. Even though each of these methods represents the exposure of the monolayer to different conditions, e.g., temperature, reactive species and deposition rate (see supplementary information S1 for more details), we are able to comprehensively quantify their effect on the monolayer in terms of the increased charge doping levels and defect densities, by analysing both the PL emission and Raman scattering spectra. We conclude that the EBE deposition process is the least destructive one in terms of preserving the exciton emission and optical properties of the monolayer.

The monolayer $WS_2$ used in this study was obtained by using the conventional mechanical exfoliation technique [20] from the commercial bulk $WS_2$ crystal (sourced from HQ graphene [21]), then transferred on top of a commercial $SiO_2$ substrate (sourced from Nova electronic materials [22]), as it shown in Fig. 1(a). Following the transfer, 20 nm of a high-$\kappa$ dielectric ($Al_2O_3$, $\kappa \approx 9$; $SiN_x$, $\kappa \approx 7$ [23]), or a low-$\kappa$ dielectric ($SiO_2$, $\kappa \approx 3.9$ [23]) material was deposited on top of separate $WS_2$ monolayers by four different thin-film deposition techniques: ALD, PECVD, EBE, or sputtering. For technical details see Supplementary Information (S1). The structure of the encapsulated samples is shown schematically in Fig. 1(b). The as-exfoliated monolayer and the encapsulated samples were characterized with the atomic force microscope (AFM) (see Supplementary Information S2), photoluminescence, reflectivity, and Raman studies. PL and Raman spectra were probed at room temperature by exciting the samples with a frequency doubled ND:YAG continuous wave (cw) 532 nm laser source, with the energy (E ≈ 2.33 eV) above the optical bandgap of monolayer $WS_2$ [24]. The reflectivity measurements were performed with a tungsten halogen white light source.

**Results and discussion**

Figure 1(c) shows the PL spectrum, whose peak energy and intensity confirms the monolayer nature of the as-exfoliated material [24]. The PL line profile is fitted with a three-peak Voigt profile for the exciton ($X^0$), trion ($X^-$), and localized states (LS) peaks [25, 26]. The $X^0$ peak normally arises from neutral exciton recombination at the K points of the Brillouin Zone (BZ), where direct optical transition of the electrons and holes occurs, the $X^-$ peak is associated with the recombination of charged excitons, and the LS peak is related to the excitons bound to structural disorders and impurities. The fitting results place the exciton energy ($X^0$ peak) at 2.014 eV, and the trion energy ($X^-$ peak) at 1.982 eV, which is in good agreement with the previously reported data for monolayer $WS_2$ [27, 28]. Figure 1(d) shows the quasi-resonant (in resonance with B exciton absorption [29]) Raman spectra of the exfoliated flake with the deconvoluted fine peaks. Raman spectroscopy, as a non-destructive characterization method, is widely used to gain insight into the vibrational modes of the material system [30]. Due to the $D_{3h}$ point group symmetry of the monolayer TMDCs [31], nine vibration modes at the center of the BZ ($\Gamma$ point) give rise to three acoustic phonon branches and six optical phonon branches [32]. The three acoustic branches are the in-plane longitudinal (LA), in-plane transverse (TA) and out-of-plane (ZA) modes, and the six optical branches are two in-plane longitudinal ($LO_1$ and $LO_2$), two in-plane transverse ($TO_1$ and $TO_2$) and two out-of-plane ($ZO_1$ and $ZO_2$) modes. These phonon modes at the $\Gamma$ point give rise to the irreducible representations E′ ($LO_2+TO_2$), E″($LO_1+TO_1$), A′($ZO_2$) and A″($ZO_1$), where E′, E″ and A′ are Raman active modes, and A″ is infrared (IR) active [33-35]. Among the Raman active modes, E′ and A′ modes are observed at ~353 $cm^{-1}$ and ~414 $cm^{-1}$ in the Raman spectra, which are recognized as first-order Raman modes [36, 37]. However, the E″ mode at the $\Gamma$ point is not Raman active in a back-scattered configuration [37, 38], therefore it is not presented in our Raman spectra. The wavenumber difference between the E′ mode and A′ mode is around 61 $cm^{-1}$, which is a typical value for a monolayer $WS_2$ [34]. Away from the $\Gamma$ point, the dispersion of the normal vibration modes at the edge of the BZ (M point) gives rise to other Raman peaks, which are reported to have a good correlation with the density of disorder states [39]. The Raman spectrum of our sample clearly shows these small peaks, which are labelled according to the previous studies as LA(M) (~ 170 cm$^{-1}$), 2ZA(M) (~293 cm$^{-1}$), E″(M) (~321 cm$^{-1}$) and 2LA(M) (~348 cm$^{-1}$) [33, 34, 40-42].

The surface morphology of the deposited material for all encapsulated samples were characterized with the AFM measurement, and the results shown in Fig. S2 (see Supplementary Information) indicate uniform coverage of the dielectric layer. The consistent height profile of the capped monolayers (~3-4 nm) is in good agreement with the height profile of the as-exfoliated monolayer (~2.6 nm), showing that the monolayers are homogeneously covered. This value is higher than the typical thickness of monolayer WS$_2$ ~0.8 nm [43], probably due to a pronounced van der Waals gap between the dry-transferred flake and the substrate, as shown in other layered structures [44, 45].

emission of the encapsulated monolayers is quenched. The sample encapsulated by ALD of Al$_2$O$_3$ preserves the highest quantum yield, which is most likely due to the low-energy of the deposited material that is chemically grown on the material surface layer by layer [46]. This is followed by the samples encapsulated by EBE (EBE Al$_2$O$_3$ and EBE SiO$_2$) and the samples encapsulated by PECVD (PECVD SiO$_2$ and PECVD SiN$_x$), in which particles are mainly physically deposited on the monolayer surface [47]. As for the sputtered samples (both Al$_2$O$_3$ and SiO$_2$), even though the monolayers are still physically present, the exciton emission, compared to the other samples, is massively quenched and the monolayer is optically dead. This suggests that sputtering is the most aggressive and damaging dielectric deposition method for WS$_2$. The strong exciton quenching after the sputtering could be the consequence of a strong increase of the defect densities in the monolayer crystal structure caused by the high

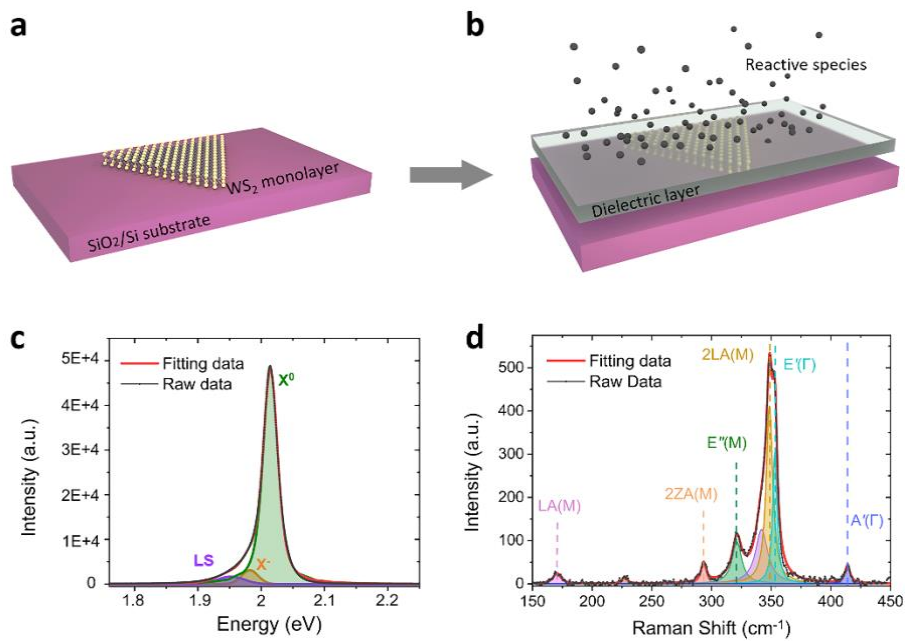

**Figure 1**. (a) Schematics of a WS$_2$ monolayer on SiO$_2$/Si substrate; (b) Schematics of a WS$_2$ monolayer encapsulated by a dielectric layer; (c) PL spectra of the as-exfoliated WS$_2$ monolayer; (d) Raman spectra of the as-exfoliated WS$_2$ monolayer. The raw data and the result of multi-peak fitting in (c,d) are shown by solid black and red lines, respectively, with the shaded areas corresponding to individual peaks.

To study the effect of the different dielectric deposition processes on the optical properties of the monolayer, we probed the PL and the reflectivity spectra for the encapsulated samples and compared them to the non-encapsulated (as-exfoliated) monolayer WS$_2$. Figure S3 in Supplementary Information contains the PL intensity maps for all samples, showing the homogeneous PL textures for the bare monolayer and encapsulated samples with different intensities, which implies that the influence of the dielectric encapsulation on the optical properties is uniform over the monolayer WS$_2$. Figure 2(a) shows the corresponding PL spectra for all samples, averaged over a certain sample area. The overall PL intensities, and, in particular, the exciton kinetic energy particles bombarding the monolayers during the sputtering process.

The reflectance contrast spectra of the samples, $\Delta R/R_{\text{ref}} = (R - R_{\text{ref}})/R_{\text{ref}}$, where $R_{\text{ref}}$ is the reflectance of the substrate, are shown in Fig. S4 (see Supplementary Information). Their derivatives with respect to the excitation energy are presented in Fig. 2(b) and highlight the absorption features. These results allow us to compare the exciton oscillator strength in these samples, which quantifies the interaction strength between excitons and light [48]. In particular, the oscillator strength of the excitons is proportional to the product of linewidth and amplitude of the reflectance dip (absorption peak) at the exciton resonance [49]. The results in Fig. 2(b) indicate that all dielectric deposition processes reduce the exciton oscillator strength. The samples encapsulated by EBE $Al_2O_3$ and EBE $SiO_2$ maintain the highest oscillator strength, followed by the ALD $Al_2O_3$ sample and PECVD samples. Furthermore, ALD $Al_2O_3$ and PECVD $SiN_x$ spectra display a non-negligible absorption feature emerging at the trion energy (highlighted in the curve). This indicates that the charge doping effect has a significant influence on the optical properties in these samples. The quenching of both exciton emission and oscillator strength suggests that the material quality is degraded after the direct deposition of a dielectric, which can be due to the introduced disorders, doping and even progressive material ablation. The influence of different factors is further analysed in the following sections. Interestingly, the monolayer encapsulated by ALD $Al_2O_3$ has a low exciton oscillator strength, but the highest PL emission. This might be due to the successive doping during the ALD process, which consists of a series of alternating precursor and purging gas injections in an ALD reaction chamber, whereby the deposition of thin films relies on the chemical reactions between the precursors directly on top of the material surface. Therefore, as a consequence of the chemical reaction effect, the band structure and optical response of the monolayer might change significantly. A quantitative analysis of this effect is subject to further studies. In contrast, the other deposition techniques mainly rely on the vapour that is initiated physically or chemically. Therefore, the monolayer properties are not affected by the direct chemical reactions at the interface during the deposition process.

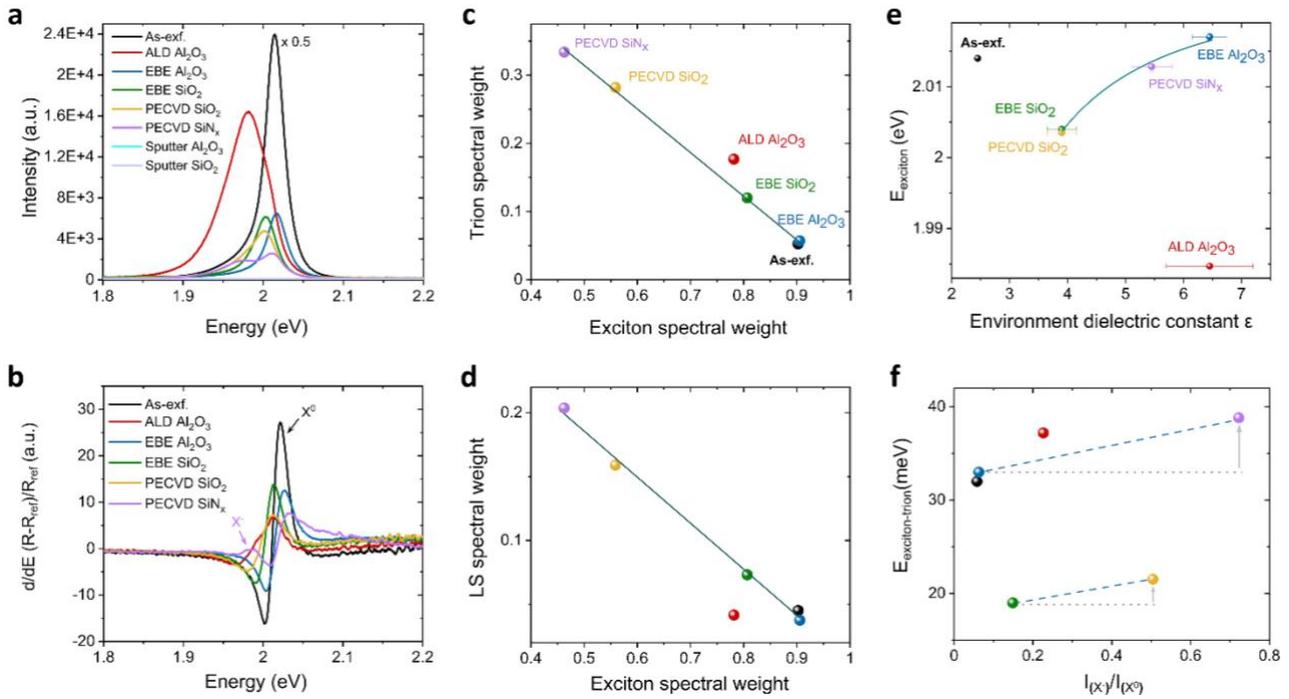

**Figure 2**. (a) Quantitative comparison of the PL spectra of the as-exfoliated monolayer $WS_2$ and the monolayers encapsulated by a dielectric material via ALD, EBE, PECVD, or sputtering; (b) Derivative of the reflectance contrast spectra of the as-exfoliated monolayer $WS_2$ and the encapsulated monolayers; (c-d) Correlation between the spectral weights of the excitons and trions (c) and the excitons and the localized states (d) of the PL spectra in (a) extracted from the multi-peak fitting. Solid lines in (c,d) are linear fits to the data points excluding that of ALD $Al_2O_3$; (e) Exciton energies of the monolayers in the different dielectric environments. Horizontal error bars indicate variation in the reported values of the dielectric constants [58-66]; (f) Energy difference between the excitons and trions of the as-exfoliated monolayer $WS_2$ and the PECVD and EBE encapsulated monolayers, the grey arrows mark the doping dependent energy splitting of the low-$\kappa$ and high-$\kappa$ capped samples, respectively. Due to the negligible optical response and the extremely low PL signal of the samples after sputtering, the reflectivity spectra and the fitting results for this method are not shown in (b-f).

The underlying peaks in the PL spectra presented in Fig. 2(a) are extracted by using the same fitting method as the one employed in Fig. 1(b). The fitting results shown in Figs. 2(c-d) demonstrate the variation in spectral contributions of the individual peaks to the PL of the different encapsulated samples. Compared to the as-exfoliated sample, the encapsulated samples suffer the reduction of exciton emission, accompanied by the rise of the trion ($X^-$) and the localised states (LS) emissions. In particular, the decline of the exciton emission shows a strong linear correlation with the increase of trion and LS emission for the EBE and PECVD samples, as opposed to the ALD sample. This indicates that, in the EBE and PECVD samples, the trion formation correlates to the density of defect states or disorders (e.g., the strain doping effect [50]), which is further demonstrated in Fig. S5 (see Supplementary

Information) showing a strong linear relationship between doping level ($X^-/X^0$) and LS spectral weight. The fact that the EBE and PECVD data points fall on the same line strongly suggests that these deposition techniques produce the same effect of different strength, with the PECVD producing more defect states and disorder. For the ALD $Al_2O_3$ sample, on the other hand, the trion emission is significant while the LS emission is minor and has the same level as that of the EBE $Al_2O_3$ sample. The doping level reflected in a significant weight of the trion emission in this case could arise from the chemical reactions during the deposition process, leading to charge transfer and chemical doping, instead of doping induced by structural defects. Moreover, EBE of $Al_2O_3$ produces a distribution of spectral weights that is the closest to that of the as-exfoliated sample, with a prominent exciton spectral weight. This indicates that the EBE deposition process results in the lowest level of doping and defect densities.

in-plane strain via lattice mismatch, etc. [55-57]. Therefore, it can be potentially affected by the deposition processes. In the ideal case of the effective band gap not shifted by the deposition process, a dielectric cap on top of a monolayer should increase the exciton energy due to reduction of the binding energy. Figure 2 (e) shows the exciton energies from the PL measurements on the different samples plotted against the external dielectric constants, which have a large range of reported values [58-66]. Here, we observe strong variation in the exciton energies of the encapsulated monolayers after the three encapsulation techniques. For the samples encapsulated by EBE and PECVD, the shifts in the exciton energy can be well described by the reported $E \sim \varepsilon^{-2}$ correlation [see Fig. 2(e)] for changes in the dielectric environment [67]. This correlation indicates that the EBE and PECVD processes induce a similar modification of the band structure and reduce the band gap compared to the as-

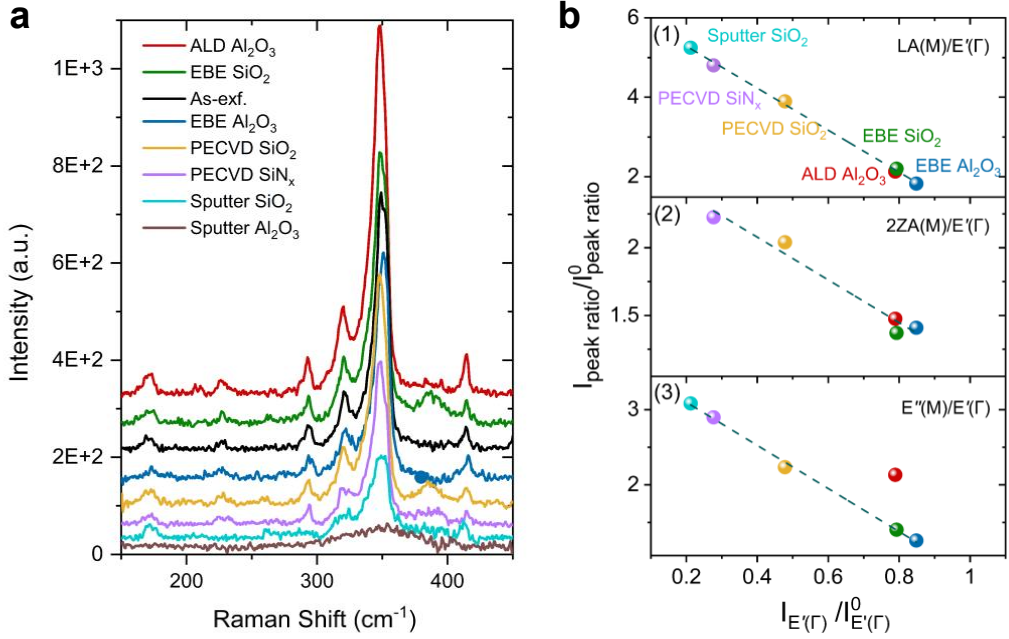

**Figure 3.** (a) Quantitative comparison of the Raman spectra of the as-exfoliated monolayer $WS_2$ and the monolayers encapsulated by ALD, EBE, PECVD, or sputtering; (b) Change of the relative intensities of the Raman peaks: LA(M) (1), 2ZA(M) (2), E″(M) (3) for the dielectric encapsulated monolayers as function of E'(Γ) peak relative to the as-exfoliated sample.

Additionally, we observe energy shifts of the excitons and trions relative to the as-exfoliated sample. The exciton transition energy is given by $E_g - E_{bX}$, where $E_g$ is the quasiparticle gap and $E_{bX}$ is the exciton binding energy [51, 52]. The exciton binding energy is inversely proportional to the square of the external dielectric constant $\varepsilon = \frac{1}{2}(\kappa_{top} + \kappa_{bottom})$ due to the dielectric screening between the charge carriers [51, 53, 54], where $\kappa_{top}$ and $\kappa_{bottom}$ are the dielectric constant of the top and bottom dielectric layer, respectively. On the other hand, the band gap can be effectively shifted by modifying the material system, e.g., by varying the carrier densities, lattice constants via mechanical strain, exfoliated monolayer. This agrees well with the strong linear relationship between the two methods presented in Fig. 2(c-d). Furthermore, for the same technique, one can easily observe the expected reduction in the exciton binding energy as the dielectric constant increases. In particular, a ~13 meV blueshift is observed from $SiO_2$ to $Al_2O_3$ for EBE, and a ~9 meV blueshift from $SiO_2$ to $SiN_x$ for PECVD. However, the sample capped by ALD of $Al_2O_3$ does not follow this trend and has a large redshift (~29 meV) compared to the as-exfoliated monolayer [19]. Compared to the EBE and PECVD samples, the bandgap after the ALD deposition process is additionally reduced by around 32 meV. This is most likely due to the chemical processes involved in this method, therefore this sample is not directly comparable to the samples capped by EBE and PECVD.

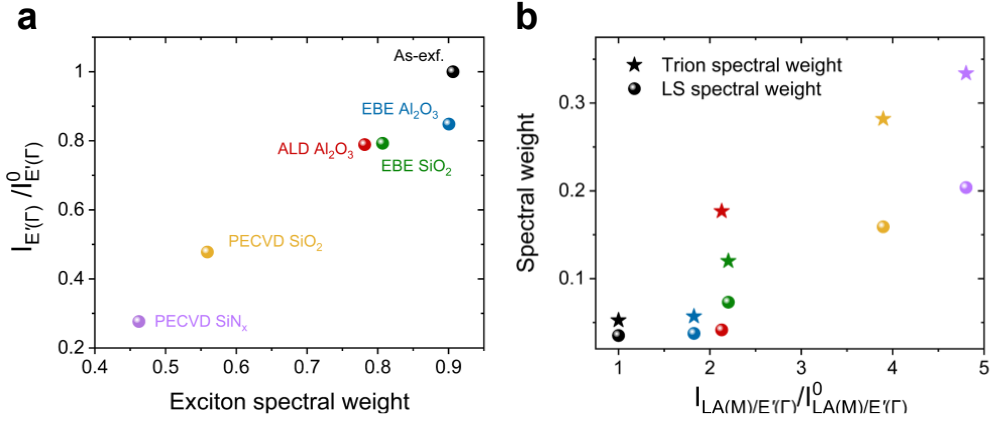

**Figure 4.** (a) Correlation between the E'(Γ) mode (normalized to the as-exfoliated sample) and the exciton spectral weight; (b) Correlation between the LA(M) relative to the E'(Γ) mode (normalized to the as-exfoliated sample) and the trion and LS spectral weight

The difference between the exciton and trion energies, presented in Fig. 2(f), shows strong dependence on the doping level, which increases with the trion spectral weight and the dielectric constant. Because ALD of $Al_2O_3$ has a very different effect on the $WS_2$ band structure compared to the PECVD and EBE processes, as discussed above, it is excluded from this comparison. The energy splitting differs when the materials with the same (or similar) dielectric constant are deposited via different methods. Namely, the sample encapsulated by PECVD of $SiO_2$ has a larger exciton-trion splitting than the sample encapsulated by EBE of $SiO_2$, and the sample encapsulated by PECVD of $SiN_x$ has a larger exciton-trion splitting than the sample encapsulated by EBE of $Al_2O_3$, as shown by the pale grey arrows in Fig. 2(f). This variation in the energy difference could arise due to the increased n-type doping level caused by PECVD. Electrons dressing the excitons give rise to polaron effects, as observed in a gated TMDC sample [68]. This effect could lead to the apparent "repulsion" of the exciton and trion levels observed at a higher electron density, which can explain the larger energy splitting observed in our samples with higher level of doping.

In order to further investigate the structural change of the monolayer lattice caused by the dielectric deposition processes, we performed a quasi-resonant Raman scattering measurement. Figure 3(a) shows representative Raman spectra for all the encapsulated samples and the as-exfoliated monolayer. The changes of different Raman peaks are analysed by fitting the Raman spectra using the same fitting method as in Fig. 1(c). We do not observe any significant peak shifts after either of the four deposition processes, which indicates that the capped monolayers are not severely affected by the external perturbation, such as the uniaxial strain effect [69]. However, the softening and broadening of the E'(Γ) (~353 cm$^{-1}$) peak, which is activated by the in-plane displacement of W and S atoms, is strongly pronounced in the samples encapsulated by sputtering of $Al_2O_3$ and $SiO_2$, followed by the PECVD samples, while the EBE and ALD samples are only weakly affected. The activation of Raman modes fundamentally depends on the crystallinity of the material [32], hence the suppression of the E'(Γ) mode indicates the decrease of the domain size for nm-sized disorders [42]. This is likely caused by the hot ions bombarding the monolayer during the fabrication process [39], which might even break the lattice symmetry [69]. This further confirms that sputtering is the most aggressive of all investigated deposition techniques. Moreover, we find the correlation between the zone-edge (M point) modes with the zone-center (Γ point) modes, as shown in Fig.3(b). Here, the fitting results for the sample capped with sputtered $Al_2O_3$ are not included since the corresponding Raman features are not present in the spectrum. The peak intensity ratios are shown with respect to the E'(Γ) peak intensity, which are normalized relative to that of the as-exfoliated monolayer. The Raman peaks that are associated with the zone-edge phonon modes are likely to be activated by the momentum contribution from localized defects which can help to satisfy the Raman selection rule [36]. The correlation between these Raman peaks and defects has been studied in different TMDC flakes [70-72], with LA(M) mode (~ 170 cm$^{-1}$), arising from the LA acoustic phonon branch at the M point, shown to be the most prominent peak [39]. Figure 3(b)-(1) shows a strong linear correlation of LA(M) and E'(Γ) with a negative slope, which implies that the decreasing crystallinity of the monolayer is directly related to the increase of defect densities, and the structural disorder and localized defects are simultaneously introduced during the deposition processes. Furthermore, 2ZA(M) (~293 cm$^{-1}$) and 2LA(M) (~348 cm$^{-1}$) modes are observed, which originate from the scattering by a combination of two ZA phonons and two LA phonons in the proximity of the M point, respectively [36]. The 2LA(M) mode is affected by the band structure due to the intervalley scattering, according to the previous studies [42, 73].

However, given the separation of only ~5 cm$^{-1}$ from the E′(Γ) mode, evaluating the intensity distribution between these two Raman modes and extracting the absolute intensity of the 2LA(M) mode from the Raman spectra cannot be done precisely [73]. Therefore, the changes in the 2LA(M) mode intensities are not analysed in our study. As for the 2ZA(M) mode, its intensity shows the inverse correlation to the E′(Γ) mode [Fig. 3(b)-(2)], similar to that of the LA(M) mode (sputtered $SiO_2$ is not present since the Raman mode is not present in the spectrum). This indicates that the 2ZA(M) mode is also affected by the defect densities.

Additionally, the peak at 321 cm$^{-1}$ is assigned as E″(M) mode and originates from the $TO_1$ phonon dispersion at the edge of the BZ [32, 33]. Although E″ process is normally Raman inactive in the back-scattering configuration [38], due to the excitonic resonant effect, the observation of this mode has been reported in the resonant Raman scattering in different TMDCs flakes [74-76]. The deviation of the E″(M) mode, which is shown in Fig. 3(b)-(3), shows a negative linear correlation with the E′(Γ) mode for the samples encapsulated by sputtering, PECVD and EBE, while the ALD sample does not follow this trend. This implies that the E″(M) mode might be affected by the interplay between defects and optical transitions at the K point of the BZ. Since the optical response for the sample encapsulated by ALD is additionally affected by the chemical reactions compared to the other deposition techniques, as discussed in the PL analysis above, the ALD sample deviates from the linear correlation between the E′(Γ) and the E″(M) mode intensities present in the other samples. Overall, it is clear that the monolayer encapsulated by EBE $Al_2O_3$ maintains the highest crystal quality and is least affected by disorder.

Finally, we investigate the correlations between the PL and Raman spectra for the samples in which the exciton PL could be measured. Figure 4(a) shows a good correlation between the E′(Γ) mode peak intensity and the exciton spectral weight, which links the decreasing contribution of exciton to the optical response with the deterioration of crystal quality. This further confirms that the quenching of the excitons in the capped samples is caused by disorder and defects. Moreover, Fig. 4(b) shows the correlation between the LA(M) peak intensity and the spectral weights of the excitonic complexes, i.e., trions and the LS. This result suggests that the different types of defects that activate the LA(M) mode promote the formation of trions and bound excitons in the samples encapsulated by the techniques that are mainly driven by physical rather than chemical process (EBE and PECVD). However, while the defects activating the LA(M) mode cause a similar level of LS formation in the ALD $Al_2O_3$ sample compared to the EBE $Al_2O_3$ sample, the trion formation is strongly enhanced. This highlights that the ALD is very different to the other deposition processes and implies that the strong doping after ALD of $Al_2O_3$ is more likely due to chemical reactions during the deposition process rather than the formation of defects (vacancies) that promote the formation of bound excitons. Overall, these results suggest that all direct deposition techniques used in this work create defects, which leads either to increased doping, or to the formation of localized states, or both, subsequently reducing the yield of the neutral excitons.

In summary, we explored the effect of different dielectric deposition techniques on the optical response of semiconducting monolayer $WS_2$. The reduction of the PL emission intensity results from the exciton quenching by all fabrication method. By comparing the doping level and emission from the localized states, we conclude that EBE is the least aggressive deposition method, followed by ALD, PECVD, and magnetron sputtering. The changes in the exciton energies arise from band gap shifts after the deposition processes and dielectric screening induced shifts of the exciton binding energies. By comparing the exciton energies, we show that all processes effectively reduce the quasi-particle band gap, which has a similar magnitude after the EBE and PECVD processes. The Raman spectra of the samples after the dielectric deposition reveal quenching of the E′ peak that is highly correlated to the quenching of the exciton PL, suggesting the strong role of disorder in exciton quenching. The comparison of Raman peaks associated with zone-edge related, defect-activated modes reveal that encapsulation by EBE of $Al_2O_3$ is the least destructive deposition method, and that sputtering of $Al_2O_3$ causes the greatest damage to the monolayer. We also show that the trion and localized exciton formation directly scales with the defect-activated LA(M) Raman peaks for the EBE and PECVD samples, but not for the sample encapsulated by ALD $Al_2O_3$. Our work provides a comprehensive survey of dielectric encapsulation methods for monolayer $WS_2$, which is critically important for guiding the integration of TMDCs into functional optoelectronic devices. In addition, our results motivate further investigation and improvement of non-destructive dielectric deposition techniques for 2D semiconductors.

### Acknowledgments

This worked was supported by the Australian Research Council (ARC) through the Centre of Excellence grant CE170100039. We acknowledge the technical support of the sample fabrication from the Australian National Fabrication Facility (ANFF) at its nodes in Victoria and Australian Capital Territory (ACT).

# References

[1] Xia F, Wang H, Xiao D, Dubey M, et al. 2014 Two-dimensional material nanophotonics *Nature Photonics* **8** 899-907

[2] Duan X, Wang C, Pan A, Yu R, et al. 2015 Two-dimensional transition metal dichalcogenides as atomically thin semiconductors: opportunities and challenges *Chemical Society Reviews* **44** 8859-76

[3] Wang Q H, Kalantar-Zadeh K, Kis A, Coleman J N, et al. 2012 Electronics and optoelectronics of two-dimensional transition metal dichalcogenides *Nature Nanotechnology* **7** 699-712

[4] Chhowalla M, Shin H S, Eda G, Li L-J, et al. 2013 The chemistry of two-dimensional layered transition metal dichalcogenide nanosheets *Nature Chemistry* **5** 263-75

[5] Hong H, Liu C, Cao T, Jin C, et al. 2017 Interfacial engineering of Van der Waals coupled 2D layered materials *Advanced Materials Interfaces* **4** 1601054

[6] Das S, Chen H-Y, Penumatcha A V and Appenzeller J 2013 High performance multilayer $MoS_2$ transistors with scandium contacts *Nano Letters* **13** 100-5

[7] Dickinson R G and Pauling L 1923 The crystal structure of molybdenite *Journal of the American Chemical Society* **45** 1466-71

[8] Cong C, Shang J, Wang Y and Yu T 2018 Optical properties of 2D semiconductor $WS_2$ *Advanced Optical Materials* **6** 1700767

[9] Ross J S, Wu S, Yu H, Ghimire N J, et al. 2013 Electrical control of neutral and charged excitons in a monolayer semiconductor *Nature Communications* **4** 1474

[10] Jones A M, Yu H, Ghimire N J, Wu S, et al. 2013 Optical generation of excitonic valley coherence in monolayer $WSe_2$ *Nature Nanotechnology* **8** 634-8

[11] Wang H, Yu L, Lee Y-H, Shi Y, et al. 2012 Integrated circuits based on bilayer $MoS_2$ transistors *Nano Letters* **12** 4674-80

[12] Lee H S, Min S-W, Chang Y-G, Park M K, et al. 2012 $MoS_2$ Nanosheet phototransistors with thickness-modulated optical energy gap *Nano Letters* **12** 3695-700

[13] Pospischil A, Furchi M M and Mueller T 2014 Solar-energy conversion and light emission in an atomic monolayer $p–n$ Diode *Nature Nanotechnology* **9** 257-61

[14] Fogler M, Butov L and Novoselov K 2014 High-temperature superfluidity with indirect excitons in Van der Waals heterostructures *Nature Communications* **5** 4555

[15] Song J-G, Kim S J, Woo W J, Kim Y, et al. 2016 Effect of $Al_2O_3$ Deposition on performance of top-gated monolayer $MoS_2$-based field effect transistor *ACS Applied Materials & Interfaces* **8** 28130-5

[16] Price K M, Najmaei S, Ekuma C E, Burke R A, et al. 2019 Plasma-enhanced atomic layer deposition of $HfO_2$ on monolayer, bilayer, and trilayer $MoS_2$ for the integration of high-κ dielectrics in two-dimensional devices *ACS Applied Nano Materials* **2** 4085-94

[17] Nam T, Seo S and Kim H 2020 Atomic layer deposition of a uniform thin film on two-dimensional transition metal dichalcogenides *Journal of Vacuum Science & Technology A: Vacuum, Surfaces, and Films* **38** 030803

[18] McDonnell S, Brennan B, Azcatl A, Lu N, et al. 2013 $HfO_2$ on $MoS_2$ by atomic layer deposition: adsorption mechanisms and thickness scalability *ACS Nano* **7** 10354-61

[19] Kim S Y, Yang H I and Choi W 2018 Photoluminescence quenching in monolayer transition metal dichalcogenides by $Al_2O_3$ encapsulation *Applied Physics Letters* **113** 133104

[20] Novoselov K S, Geim A K, Morozov S V, Jiang D, et al. 2004 Electric field effect in atomically thin carbon films *Science* **306** 666-9

[21] http://www.hqgraphene.com.

[22] http://www.novawafers.com.

[23] Robertson J 2004 High dielectric constant oxides *The European Physical Journal Applied Physics* **28** 265-91

[24] Zeng H, Liu G-B, Dai J, Yan Y, et al. 2013 Optical signature of symmetry variations and spin-valley coupling in atomically thin tungsten dichalcogenides *Scientific Reports* **3** 1608

[25] Tongay S, Suh J, Ataca C, Fan W, et al. 2013 Defects Activated photoluminescence in two-dimensional semiconductors: interplay between bound, charged and free excitons *Scientific Reports* **3** 2657

[26] Chow P K, Jacobs-Gedrim R B, Gao J, Lu T-M, et al. 2015 Defect-induced photoluminescence in monolayer semiconducting transition metal dichalcogenides *ACS Nano* **9** 1520-7

[27] Shang J, Shen X, Cong C, Peimyoo N, et al. 2015 Observation of excitonic fine structure in a 2D transition-metal dichalcogenide semiconductor *ACS Nano* **9** 647-55

[28] Xu W, Kozawa D, Liu Y, Sheng Y, et al. 2018 Determining the optimized interlayer separation distance in vertical stacked 2D $WS_2$:hBN:$MoS_2$ heterostructures for exciton energy transfer *Small* **14** 1703727

[29] Zhu B, Chen X and Cui X 2015 Exciton binding energy of monolayer $WS_2$ *Scientific Reports* **5** 9218

[30] Chakraborty B, Bera A, Muthu D, Bhowmick S, et al. 2012 Symmetry-dependent phonon renormalization in monolayer $MoS_2$ transistor *Physical Review B* **85** 161403

[31] Ribeiro-Soares J, Almeida R, Barros E B, Araujo P T, et al. 2014 Group theory analysis of phonons in two-dimensional transition metal dichalcogenides *Physical Review B* **90** 115438

[32] Zhang X, Qiao X-F, Shi W, Wu J-B, et al. 2015 Phonon and Raman scattering of two-dimensional transition metal dichalcogenides from monolayer, multilayer to bulk material *Chemical Society Reviews* **44** 2757-85

[33] Molina-Sanchez A and Wirtz L 2011 Phonons in single-layer and few-layer $MoS_2$ and $WS_2$ *Physical Review B* **84** 155413

[34] Zhao W, Ghorannevis Z, Amara K K, Pang J R, et al. 2013 Lattice dynamics in mono-and few-layer sheets of $WS_2$ and $WSe_2$ *Nanoscale* **5** 9677-83

[35] Ataca C, Topsakal M, Akturk E and Ciraci S 2011 A comparative study of lattice dynamics of three-and two-dimensional $MoS_2$ *The Journal of Physical Chemistry C* **115** 16354-61

[36] Gontijo R N, Resende G C, Fantini C and Carvalho B R 2019 Double resonance raman scattering process in 2D materials *Journal of Materials Research* **34** 1976-92

[37] Saito R, Tatsumi Y, Huang S, Ling X, et al. 2016 Raman spectroscopy of transition metal dichalcogenides *Journal of Physics: Condensed Matter* **28** 353002

[38] Molas M R, Nogajewski K, Potemski M and Babiński A 2017 Raman scattering excitation spectroscopy of monolayer $WS_2$ *Scientific Reports* **7** 5036

[39] Mignuzzi S, Pollard A J, Bonini N, Brennan B, et al. 2015 Effect of disorder on Raman scattering of single-layer MoS$_2$ *Physical Review B* **91** 195411

[40] Mitioglu A, Plochocka P, Deligeorgis G, Anghel S, et al. 2014 Second-order resonant raman scattering in single-layer tungsten disulfide WS$_2$ *Physical Review B* **89** 245442

[41] Berkdemir A, Gutiérrez H R, Botello-Méndez A R, Perea-López N, et al. 2013 Identification of individual and few layers of WS$_2$ using Raman spectroscopy *Scientific Reports* **3** 1755

[42] Shi W, Lin M-L, Tan Q-H, Qiao X-F, et al. 2016 Raman and photoluminescence spectra of two-dimensional nanocrystallites of monolayer WS$_2$ and WSe$_2$ *2D Materials* **3** 025016

[43] Cong C, Shang J, Wu X, Cao B, et al. 2014 Synthesis and optical properties of large-area single-crystalline 2D semiconductor WS$_2$ monolayer from chemical vapor deposition *Advanced Optical Materials* **2** 131-6

[44] Walsh L A and Hinkle C L 2017 Van der Waals epitaxy: 2D materials and topological insulators *Applied Materials Today* **9** 504-15

[45] Datta R S, Syed N, Zavabeti A, Jannat A, et al. 2020 Flexible two-dimensional indium tin oxide fabricated using a liquid metal printing technique *Nature Electronics* **3** 51-8

[46] Ritala M and Leskelä M 2002 *Handbook of Thin Films* (Elsevier) pp 103-59

[47] Khodier S and Sidki H 2001 The effect of the deposition method on the optical properties of SiO$_2$ thin films *Journal of Materials Science: Materials in Electronics* **12** 107-9

[48] Schneider C, Glazov M M, Korn T, Höfling S, et al. 2018 Two-dimensional semiconductors in the regime of strong light-matter coupling *Nature Communications* **9** 2695

[49] Lundt N, Maryński A, Cherotchenko E, Pant A, et al. 2016 Monolayered MoSe$_2$: A candidate for room temperature polaritonics *2D Materials* **4** 015006

[50] Niehues I, Schmidt R, Drüppel M, Marauhn P, et al. 2018 Strain control of exciton–phonon coupling in atomically thin semiconductors *Nano Letters* **18** 1751-7

[51] Raja A, Chaves A, Yu J, Arefe G, et al. 2017 Coulomb engineering of the bandgap and excitons in two-dimensional materials *Nature Communications* **8** 15251

[52] Chernikov A, Berkelbach T C, Hill H M, Rigosi A, et al. 2014 Exciton binding energy and nonhydrogenic rydberg series in monolayer WS$_2$ *Phys Rev Lett* **113** 076802

[53] Stier A V, Wilson N P, Clark G, Xu X, et al. 2016 Probing the influence of dielectric environment on excitons in monolayer WSe$_2$: Insight from high magnetic fields *Nano Letters* **16** 7054-60

[54] Hsu W-T, Quan J, Wang C-Y, Lu L-S, et al. 2019 Dielectric impact on exciton binding energy and quasiparticle bandgap in monolayer WS$_2$ and WSe$_2$ *2D Materials* **6** 025028

[55] Chernikov A, Ruppert C, Hill H M, Rigosi A F, et al. 2015 Population inversion and giant bandgap renormalization in atomically thin WS$_2$ layers *Nature Photonics* **9** 466-70

[56] Klingshirn C F 2012 *Semiconductor Optics* (Springer Science & Business Media)

[57] Zhao X G, Shi Z, Wang X, Zou H, et al. 2021 Band structure engineering through Van der Waals heterostructing superlattices of two-dimensional transition metal dichalcogenides *InfoMat* **3** 201-11

[58] Groner M, Fabreguette F, Elam J and George S 2004 Low-temperature Al$_2$O$_3$ atomic layer deposition *Chemistry of Materials* **16** 639-45

[59] Black L E and McIntosh K R 2012 Surface passivation of C-Si by atmospheric pressure chemical vapor deposition of Al$_2$O$_3$ *Applied Physics Letters* **100** 202107

[60] Seo M Y, Cho E N, Kim C E, Moon P, et al. 2010 *3rd International Nanoelectronics Conference (INEC)*

[61] Shamala K, Murthy L and Rao K N 2004 Studies on optical and dielectric properties of Al$_2$O$_3$ thin films prepared by electron beam evaporation and spray pyrolysis method *Materials Science and Engineering: B* **106** 269-74

[62] Vogt K, Houston M, Ceiler M, Roberts C, et al. 1995 Improvement in dielectric properties of low temperature pecvd silicon dioxide by reaction with hydrazine *Journal of Electronic Materials* **24** 751-5

[63] Yang L, Wang H, Zhang X, Li Y, et al. 2017 Thermally evaporated SiO$_2$ serving as gate dielectric in graphene field-effect transistors *IEEE Transactions on Electron Devices* **64** 1846-50

[64] Gong J, Dai R, Wang Z, Zhang C, et al. 2017 Temperature dependent optical constants for SiO$_2$ film on Si substrate by ellipsometry *Materials Research Express* **4** 085005

[65] Rahman H U, Gentle A, Gauja E and Ramer R 2008 *2008 IEEE International Multitopic Conference*

[66] Zhou X, Chen Q, Zhang Q and Zhang S 2011 Dielectric behavior of bilayer films of P(VDF-CTFE) and low temperature PECVD fabricated Si$_3$N$_4$ *IEEE Transactions on Dielectrics and Electrical Insulation* **18** 463-70

[67] Raja A, Waldecker L, Zipfel J, Cho Y, et al. 2019 Dielectric disorder in two-dimensional materials *Nature Nanotechnology* **14** 832-7

[68] Sidler M, Back P, Cotlet O, Srivastava A, et al. 2017 Fermi polaron-polaritons in charge-tunable atomically thin semiconductors *Nature Physics* **13** 255-61

[69] Zhu C, Wang G, Liu B, Marie X, et al. 2013 Strain tuning of optical emission energy and polarization in monolayer and bilayer MoS$_2$ *Physical Review B* **88** 121301

[70] Lin Z, Carvalho B R, Kahn E, Lv R, et al. 2016 Defect engineering of two-dimensional transition metal dichalcogenides *2D Materials* **3** 022002

[71] Carvalho B R, Wang Y, Mignuzzi S, Roy D, et al. 2017 Intervalley scattering by acoustic phonons in two-dimensional MoS$_2$ revealed by double-resonance Raman spectroscopy *Nature Communications* **8** 14670

[72] McCreary A, Berkdemir A, Wang J, Nguyen M A, et al. 2016 Distinct photoluminescence and raman spectroscopy signatures for identifying highly crystalline WS$_2$ monolayers produced by different growth methods *Journal of Materials Research* **31** 931-44

[73] McDevitt N, Zabinski J, Donley M and Bultman J 1994 Disorder-induced low-frequency Raman band observed in deposited MoS$_2$ films *Applied Spectroscopy* **48** 733-6

[74] Lee J-U, Park J, Son Y-W and Cheong H 2015 anomalous excitonic resonance Raman effects in few-layered MoS$_2$ *Nanoscale* **7** 3229-36

[75] Nam D, Lee J-U and Cheong H 2015 Excitation energy dependent Raman spectrum of MoSe$_2$ *Scientific Reports* **5** 17113

[76] Soubelet P, Bruchhausen A E, Fainstein A, Nogajewski K, et al. 2016 Resonance effects in the Raman scattering of monolayer and few-layer MoSe$_2$ *Physical Review B* **93** 15540